## Bibliographic information:

Outi **Alapekkala**, Juhani **Risku**

***Software startuppers took the media's paycheck. Media's fightback happens through startup culture and abstraction shifts***, ICE Conference, Trondheim Norway 2016.

This is an author-generated preprint version.

The final version is available at:

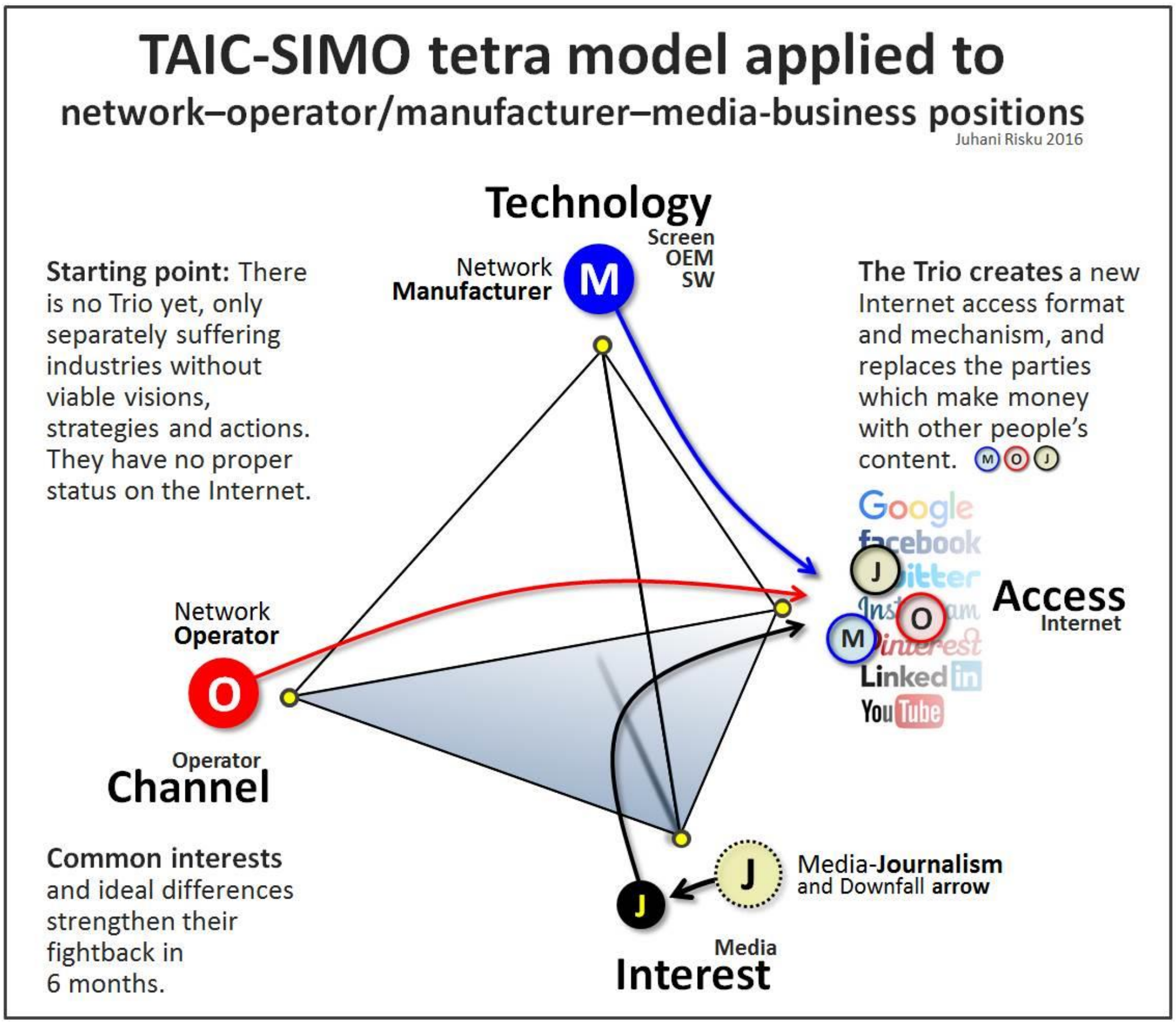

# Software startuppers took the media's paycheck

## Media's fightback happens through startup culture and abstraction shifts

Juhani Risku, Norwegian University of Science and Technology NTNU
Department of Computer and Information Science
Information Systems and Software Engineering (ISSE)
Trondheim, Norway
juhani.risku@idi.ntnu.no

Outi Alapekkala, In Action – Societal Innovation startup
Systemic functions
Media and Journalism
Tornio, Finland
outi_alapekkala@yahoo.fr

***Abstract*—The collapse of old print media and journalism happened when the Internet, its solutions, services and communities became mature and mobile devices reached the market. The reader abandoned printed dailies for free and mobile access to information. The business of core industries of the early Internet and mobile communication, the mobile network manufacturers and operators are also in stagnation and decline. Therefore these industries may have similar interests to improve or even restructure their own businesses as well as to establish totally new business models by going into media and journalism.**

**This paper analyses, first, the production flows and business models of the old and present media species. Second, it analyses the current market positioning of the network manufacturers and operators. Third, the paper suggests two avenues for media and journalism and the network manufacturers and operators, the Trio, to join their forces to update journalism and make all three stagnating industries great again. Last, we propose further research, development and discussion on the topic and envision possible futures for journalism, if the three would engage in cooperation. We see that the discussion should consist of ethical, societal and philosophical subjects because the development of the Internet solutions are based on "technology first" actions.**

**We find and outline a tremendous opportunity to create a new industry with new actors through combining the interests of the network manufacturers, network operators and journalism in a systemic solution through a strategic alliance and collaboration Fig. 1. Software startuppers with their applications and communities will be the drivers for this abstraction shift in media and journalism.**

**Our experiences in the media, journalism, mobile network, mobile phone manufacturing and startups provide the basis for our formulations on the future of those industries.**

***Keywords*—startups, media, journalism, network operators, network manufacturers, abstraction shift, creative reporter, systemic solutions, TAIC-SIMO, Cynefin**

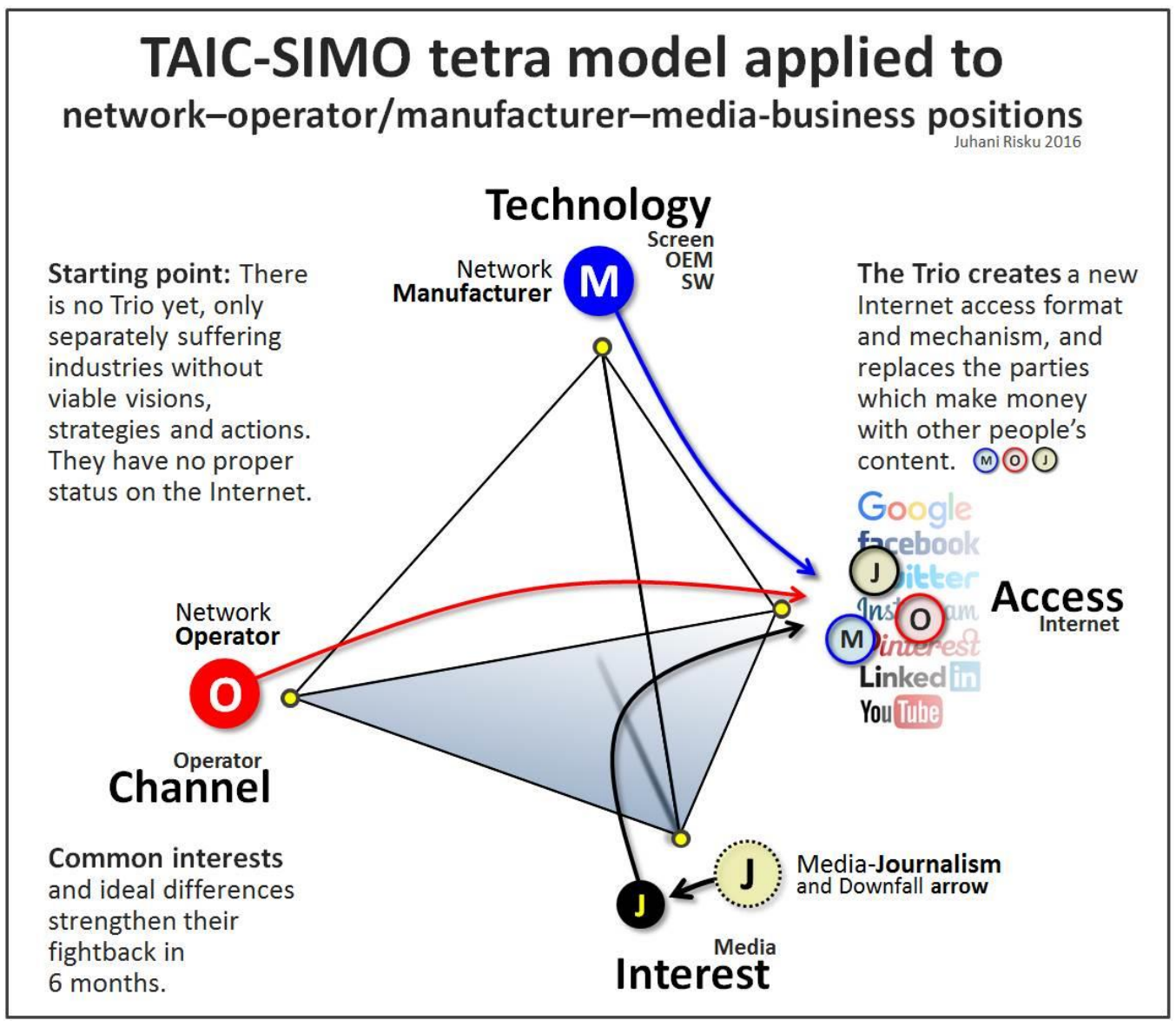

Fig. 1. TAIC-SIMO tetra model applied to network–operator/manufacturer–media-business' (the Trio's) positions. In the Trio's fightback they join their forces and build a new format and access to the Internet, and they override today's advertisement money hijackers. This requires abstraction shifts, startup culture, new leadership and rapid actions from the Trio. The tetra shows direct connections between edges (e.g. O–M collaboration), the tetra face (triangle) show a combined business area (e.g. M–O–J). The missing part of the tetra can be established on a new abstraction level (e.g. M–O–J Internet Access). This MOJ Access is a spin-off of a network–operator/manufacturer–media house consortium. The idea to establish an own Internet Access spin-off instead of acquiring an existing one is about creating a new actor in this business environment. This actor can disrupt the present business models and offer the users and consumers a combination of more interesting and fair services.

## I. Introduction

Media's and journalism's business models are today from the era of "owning the paper printing machine" and "owning the broadcast studios and channels". Owning the machine and channel was possible only for the tycoons and business moguls and they took the power through the printed newspaper and displays paid for by single copies, subscriptions and advertisements. The power came from owning the technology (expensive printing machine) and owning the channel. R.G.

Picard emphasizes that journalism's printed newspaper is extremely capital intensive business because of its high capital requirements, high fixed production, distribution, marketing and first-copy costs. [1]. Media is an umbrella term to combine business sectors and activities like broadcasting, TV, radio, Web publishing, print media and journalism into a one trade and business.

Picard underlines how journalism differs from media. Journalism is a cross-cutting discipline and a set of working methods and practices enabling content production for all publishing means from print, TV and radio to social media. He defines journalism as an activity with practices to gather, convey and process information and knowledge and insists on how journalism's functions stay prominent for society. Therefore, the practice of journalism is not a media, a distribution platform, or a business model. [2].

Having said this, journalism is in our focus to restructure the media sector business because its practices and ways of working are at the core of all media and publishing. While the media and ways to publish have evolved, no significant innovation in journalism has happened and merely "writing better articles" is clearly no longer an option. Through rethinking journalistic activity and its practices, new business can be developed that will allow media companies to overcome stagnation and to regain the customer. [3].

## II. Reasons for the lost paycheck

### *A. Startup culture replace old media's working patterns*

Software startuppers killed the traditional print journalism as they took both its readers and its income. Their algorithms have attracted the reader to access free-of-charge information and content, and advertisers have followed the customer. When software startuppers like Sergey Brin and Larry Page founded Google, it replaced older search engines like Alta Vista, Yahoo and Lycos, and made it difficult for newer engines like Ask Jeeves and Bing to gain market share. [4].

There are common factors of the media and journalism downfall: lack of innovation, comfort zone laziness, business focus failures, illusions of self-correcting actions like A. Taylor says, writing more and better articles as solution to journalism's failures [5], concentrating more on investigative journalism, improving their Web presence, publishing quality photos, visualisation. This is illusionary because journalists have certainly done good work all the time. New ideas appear include selling articles online by micro payments, and crowdsourcing as a collaborative action with the audience. A recent report commissioned to look for innovative media outlets and innovation in journalism startups found no single groundbreaking innovation. While many actions are considered as innovations in journalism, they are more of gradual improvements and tackling platforms, business models and processes, not journalistic content itself. Like Schiffrin (et.al) say, they did not find any revolutionary innovations by journalism professionals. [6].

Software startuppers' algorithms allow the theft and gathering of media outlet's public content, as well as lean and agile content production for bloggers and other citizen journalists: their articles, posts, videos and other content is published as it is ready and edited if needed. The content costs nothing to produce and nothing to publish, and it can be interactive. They write content iteratively so that articles gain readers, like Blank and Patenaude–Gaudet say. [7], [8]. In this sense, software startuppers have allowed just about anybody to become a one-man media house or a citizen journalist without having to follow some binding editorial guidelines. Membership of official professional bodies, the related fees, administrative red tape and other gate keeping activities have also become redundant.

Old media houses and journalists seek to be lean and agile through following the trend of providing part of their content for free and allowing some limited interactivity subject to moderation on their websites. They also seem to seek to attract free workforce and content through either hosting a blog platform or providing famous people from politics and business their own regular blog space.

Meanwhile software startuppers have developed their own Web presence solutions like search engines, social media chat services and short message services because they had to find business models to monetarise their companies. [9]. They harmfully put advertisement banners and later sensitive and contextual advertisement solutions on their Web pages and services.

The advertisers began to move from printed newspapers to Web services, because their visibility was nationwide or global, fees were priced by actual views of the advertisements, and the (automated) pay-per-click model (PPC) is cost efficient because of flat-rate agreements or bid-based systems. Contextual advertising programs with algorithms like Google AdWords and AdSense, and Microsoft AdCenter changed the Web advertising revolutionary from year 2006 onwards, according to Shatnawi [10]. At the same time investments in printing press, ink, paper and labour sunk. The decline of classified advertising in newspapers caused advertising revenue losses because of specialized digital online job recruitment, dating and real estate web services starting from Craigslist 1995, Leurdijk (et.al). [11].

The newspaper size format modification from broadsheet to tabloid is an indicator of print media's change: when paper consumption is halved, the number of articles has diminished by one-third. A change in the average size of articles has also occurred: there are fewer small and mid-size articles, but more large articles, like Andersson says [12]. Readers also move to superficial and sensational articles. Journalism, both in printed and investigative form is too slow in the middle of 24/7 Web publishing. Social media and new forms of instant messaging produce enormous amount of information, news and just-on-time text, while shared and combined editing offices partly cause generic offering so that the readers bump into exactly the same material on several newspapers and their Web sites. Free content becomes normality in journalism, and the original quality standards of journalism become outdated.

The rapid smartphone development started from Apple's iPhone year 2007 to accelerate the mobile revolution in mobile content creation, usage and mobile presence on news pages, knowledge search and social media. In 30 months Apple sold 42 million iPhones. As success factors for iPhone Laugesen (et.al.) mention market size, share and growth, average revenue per user (ARPU), usage of mobile data, content and services offered on AppStore, and consumer satisfaction, which in the first beginning was high. [13]. Advertisements became a part of Web and mobile content so that its share of ad turnover grew accordingly with the fall of ad turnover of the old media. It began to be difficult to make money with news and journalism, because less people were buying. [14].

*B. Startup culture redefines quality?*

As bloggers' and other citizen journalists' articles, posts and other content attract ever increasing numbers of readers and followers, old media seeks to, somewhat, undermine the quality and integrity of that content. It suggests that citizen journalism is not of as high quality, and thus not as credible as, the traditional media, which has an established profession, ethical guidelines and other general rules and norms for presenting things. For the old media quality is also a brand issue, it is a reputation built over the years on the assumption of being a trustworthy source of information.

Meanwhile, software startuppers, their algorithms, new communication platforms and applications allow the broadcasting of many more additional viewpoints, insights from professionals, experts and other stakeholders as well as for the expression of opinions that might otherwise be censored. Software startuppers have thus given platforms and potential visibility to far more views and opinions than the traditional media could have ever given. Far too strange opinions to the commonly accepted as well as stand-alone comments and insights have always been filtered out by the old media in the name of speedy production of news on all possible topics by a classic daily. A report by Johanna Vehkoo on quality journalism notes that editors and journalists basically have their own quality criteria and most publications have their own ethical code [15].

Old media no longer has the resources to do quality. Risto Uimonen suggests that editorial work has got a somewhat automated feel as the Internet values quantity and speed over quality, depth and analysis [16]. While facts and quotes may well be checked and certain ethical and journalistic standards respected, there is no time or space to voice all views. Therefore old media satisfies itself to repeat a standard explanation of events and often focuses on communicating political differences on the topics.

Citizen journalism allowed for by software startuppers upgrades the notion of freedom of speech and potential outreach of even singular opinions by allowing free publication and dissemination of one's content.

*C. Startup culture to claim the role of the Fourth Estate?*

Are press, media and journalists the Fourth Estate – the fourth power next to legislative, executive and judiciary powers, keeping a watchful eye on the three others running democratically?

Press and media affect decision-making and general opinion through deciding who and what gets visibility. In allocating this visibility, the old media sticks to old habits: reporting on and giving visibility to the views of established societal actors: governments, ministries, institutions, political parties and their people. New entrants seeking to get their voice heard are often simply dismissed or, worse, ridiculed, by the very same press and media that claim to value variety of opinions and views.

Old media is thus not the watchdog of the system as it claims, but rather the clue and visible network keeping it together and sealing the system from outsiders and new entrants.

The Internet, blogs and social media allow new entrants on the political and societal markets to voice and spread new ideas.

If the current three powers need a watchdog, the role could well be assumed by a large and democratic social media network of citizen journalists. Social media's business model is based on free speech and on the passion of professionals and experts to contribute and participate into a meaningful debate, whereas the old media cannot necessarily say and do everything because it has to keep advertisers and shareholders on board. Profoundly, journalists and media houses have different core interest: media house wants to make money for its shareholders, whereas a journalist is driven by the ideology of having a role of a societal watchdog. In the midst of these conflicting interests the weaker participant, the journalist, has to give up. This can be seen in mergers and close-downs of newspapers and magazines as well as in huge lay-offs of journalists, by Persefoni. [17]. This means that the journalists' professional ethics is diluted in the watchdog-Fourth-Estate and it is replaced by business interests of digital content services and other more popular channels for expression like chats on social media.

*D. Lean and agile reveal old media's imperfection*

Old media houses and their journalists want readers to pay for the content, events and decisions they decide to highlight or for advertisers to pay for their choices in exchange of visibility. But the reader no longer sees the point in paying for those highlights as she gets the equivalent information for free from Internet sources. Readers leave old media and advertisers follow the reader giving their money to software startuppers who have made available the multiple free platforms and services.

Software startuppers have revealed the old print media's and journalism's inadequacy, insufficiency and imperfection: old media's customer disappeared as soon as alternative sources of information become easily accessible.

In particular *four inadequacies* rise.

*One size fits all* is no longer an option. A reader does not want to buy a full paper that has information she's not interested in. She prefers going online and reading, eventually also paying for, only what she is interested in. New online platforms provide for an opportunity to establish communities defined by a common interest or topic. These social media communities and those who run them can also make money with their specialised platforms because advertisers find an extremely well targeted audience for their products and services on them. Individual bloggers who demonstrate a good number of followers are also subject to attract advertising and sponsoring money. These types of business models are of interesting value for advertisers' money and give readers direct access to topics of their interest.

*Too general and too neutral articles* of the old media are less attractive for the reader than some colourful and opinionated citizen journalist articles and a long list of readers' comments following it. Established journalists are mere observers reporting on what has been said, decided and done. They quote others sticking to what's being communicated in official press releases and by the various spokespersons and reprint information from press agencies without putting it into national and local context. Citizen journalists, bloggers and those creating content on social media platforms are themselves actors in society, doers and professionals with true insight on their topic. They are willing to contribute to societal discussion with their insights and sometimes even extremely opinionated views and other strong statements, that do not need to respect the political correctness traditional media does, in order to keep its advertisers, shareholders and access to official briefings.

*One-way communication* of the old print media is also no longer an option. The various online platforms and social media have the capacity to engage and sustain debate the old media has no capacity, resources or willingness to do. Engaging in a debate with your readers is simply not the old way of doing journalism. Readers' comments, additional information, views, opinions and corrections on content force old media to see that they don't know everything and may even have misunderstood something. Meanwhile, this agile, lean, humble and realistic attitude is fundamental for passionate citizen journalists.

*All words no deeds*. Old media is mere communication of observed events and quotes from actors. It presents the news, events and decisions as inevitable facts. "This is how it is and there's nothing you can do about it, but stay informed." Meanwhile, if it claims the role of a watchdog of other powers, it should be more engaging and suggest alternative or corrective paths and actions when it sees an injustice or an error in the system and become an actor that engages the reader. Online platforms and social media networks have a tremendous potential to initiate concrete, in particular collective, action and engage their readers in it through quick social media networking.

## III. Network Manufacturers and Operators are Strategic Relatives to the Media

Network manufacturers are few and their business is globally a mixture of expected strong growth and zero sum game. The sales of 2G and 3G networks markets will diminish steeply [18]. Growth comes from e.g. India and through 5G technologies after the year 2020. There are uncertainty factors from operators' investments in recent spectrum auctions and hard competition between network operators, says Kahn. [19]. In order to maintain their positions in the markets, mergers and acquisitions are a necessity.

The network operator industry is also under constant change Fig 2.

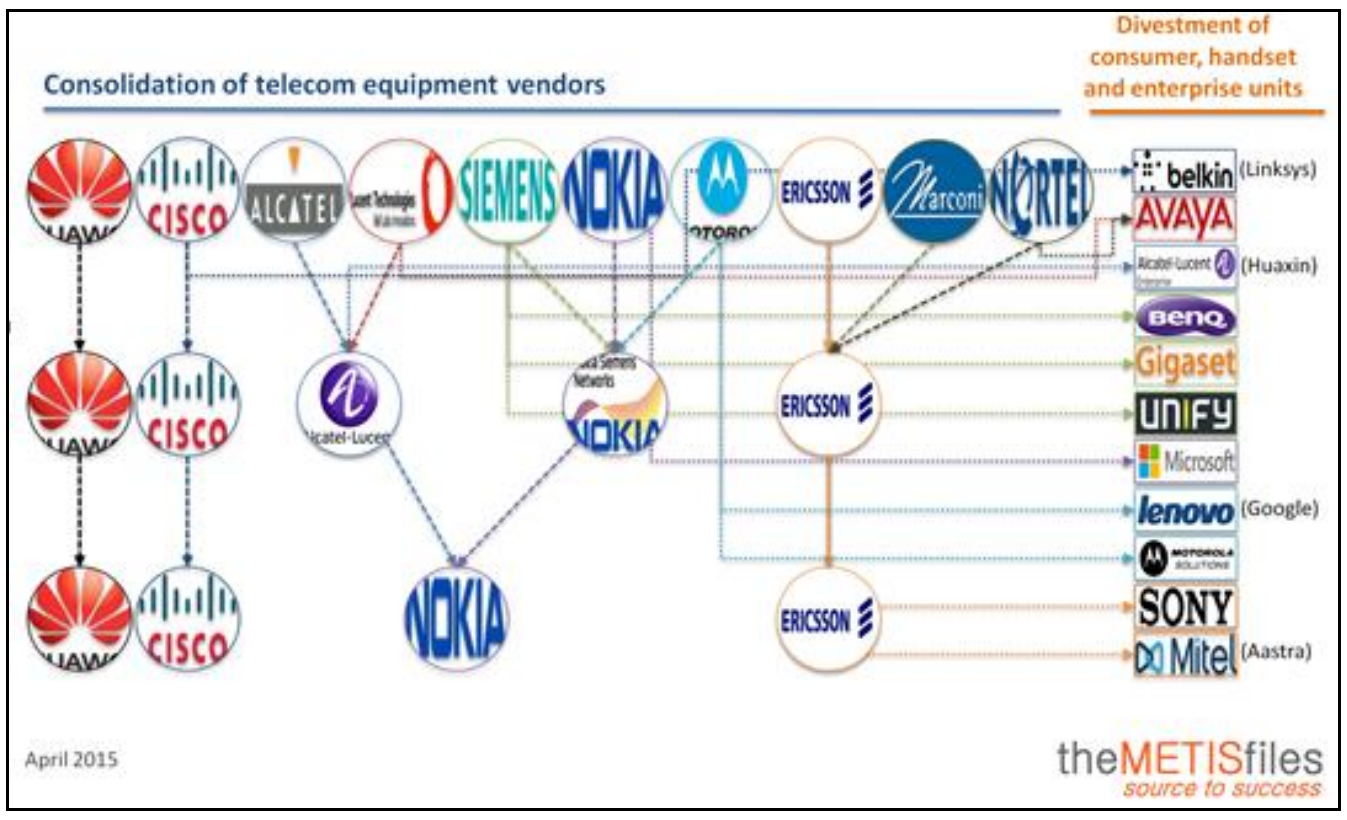

Fig. 2. How Apple, Cisco and Huawei disrupted the telecom equipment market, the Metisfiles, April 2015. [20]

The Verizon-Vodafone acquisition in 2014 was a transaction valued at approximately $130 billion. [21]. Good example of network operators' hard battle is the French companies Iliad-Free and Altice-Numericable-SFR to expand abroad and grow faster than the rival. Operators like T-Mobile US, Bouygues Télécom, Orange Suisse and Portugal Telecom have been on their shopping list. [22].

The changes in Network manufacturing and network operating businesses are as fundamental as that of journalism's. It is a battle of footprint, growth and existence.

## IV. Two Avenues and Some Winding Roads for Media-Journalism Through Alliances or Startups

When trying to give structure to operator-manufacturer-media houses', the Trio's, present position in their respective businesses, the Cynefin sense-making framework positions each of them clearly [23]. Cynefin also describes their common futures and strategic opportunities.

The Cynefin model of operator-manufacturer-media houses [Fig. 3] shows two different layers, the business awareness inside the industry (where we are today), and the ideal positioning for agile and creative future of the business (where we should be).

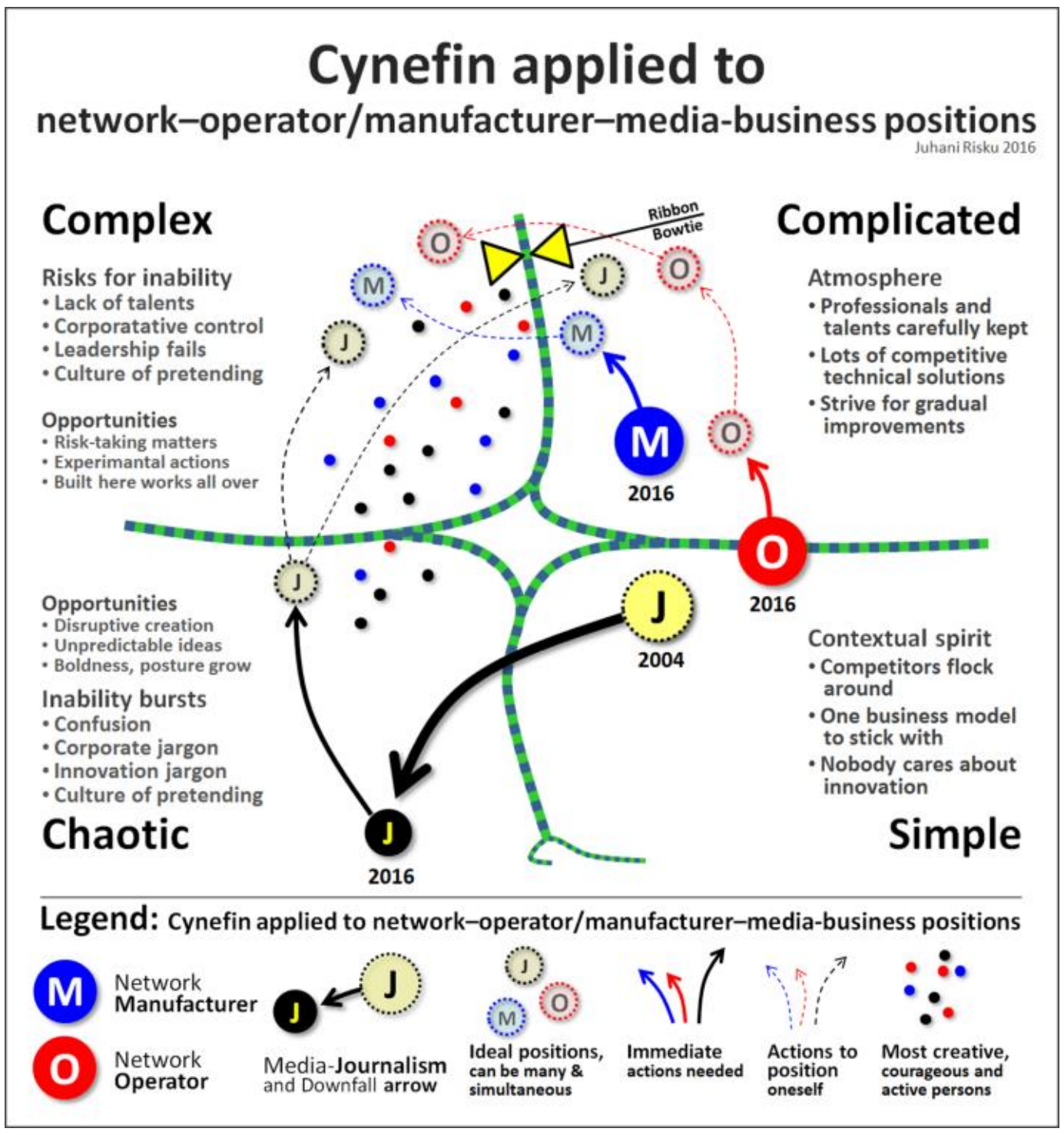

Fig. 3. Cynefin applied to network–operator/manufacturer–media-business positions 2016. Cynefin [ˈkʌnɪvɪn] [23] is framework with five domains: simple, complicated, complex, chaotic and disorder (in the intersection). Simple and complicated represent ordered world, complex and chaotic represent disordered world. When a business is in the simple section it is stable and predictable until disruptive changes happen. Companies with too high comfort zone face problems and may fall to chaotic section. The route clockwise from simple through chaos to complex and complicated is long and painstaking. Balancing in sections complex-complicated-simple allows creation, innovation and possibilities to develop stable simple businesses in a constant flow. Journalism fell from simple to chaotic, operators have stagnated more to simple, manufacturers are on the way to complex with 5G. They all need immediate actions to survive: journalism through complex to complicated, operators upwards between complicated and complex, manufacturers more to complex. Complex offers innovation and experiments, complicated offers stable business. The best place is the Ribbon-Bowtie, where good leadership guarantees innovation and successful business. The Ribbon-Bowtie positioning as a strategic endeavour requires balancing between creative and stable mindsets and giving more authority for creative and disruptive forces and talents. When M, O and J join their forces, it is a combination of creativity, making business from chaos, functioning experiments, and running solid and stable business. M, O and J have some overlap to understand each other, and they have extreme strengths to be integrated. The chaotic journey of journalism appers to be the disruptive strength of the Trio, because in its agony the industry is forced to find viable and competitive yet revolutionary solutions to survive. Now it seems that journalism has been separated from the successful media industry and it tries to survive alone.

Journalism and media houses have some ways to make steps or leaps for better business. Steps are gradual and mainly conservative, leaps are radical and innovative. We divide them to singular solutions and systemic solutions. In singular ones companies look for evolutionary steps and easy-to-communicate actions so that the stakeholders support them. Typically timid decisions like lay-offs, acquisitions, buying a startup or changing the CEO are more cosmetic actions than preparing winning businesses. In singular solutions only gradual growth and changes are allowed for understandable reasons: they are easy to communicate and to do for a corporation than turn the ship at once. It takes several years to rewrite the strategy, recruit and teach new talents, get production lines in order and get the customer to believe in the newborn brand promise. In singular solutions the superficial promise is as important as it is to look convincing. Gradual steps destroy startup culture: Skype was acquired by Microsoft, and it became part of a corporation and lost its startup culture. Nest was acquired by Google, and now Nest is losing its leaders and best people. [24]. Nest as a startup is gone.

Several corporations try to build startup culture inside the company as known as internal startups. For startuppers it may sound like adult caretakers want some entertainment from young radicals. It is impossible to think that a group of startuppers could make anything radical with limited authorisation and resources. Startuppers in a corporation are not given the role they should.

A corporation may also think that the top management and the staff can be surrounded by startup culture. A reasonable question is, that why would they become startup culture lovers by dictation and because it is fashionable? We don't have too many success stories of startup surgery in corporations.

Systemic solutions are more complex and fundamental than singular solutions. In a systemic solution a corporation or an industry envisions, plans and restructures its businesses into a new position with all available means and tools. In media, network manufacturing and network operating fields some companies could gain profits and footprint by letting an internal organization innovate and collaborate with similar organisations of other industries. This means that first steps in collaboration could be organised rather between larger teams or groups than to let only some individuals to from these companies to work together. Here these programs should be funded independently and led by someone else than the existing executives like an innovation officer or newly hired created person. Independence, freedom and changes in leadership are a crucial part of startup culture and internal startups.

Another systemic solution is to establish joint programs and strategic alliances inside one's own industry. This is lighter than a merger and acquisition, but it gives more critical mass. However, a joint program with a competitor in a stagnating business may look more like losers join their forces.

Instead, a radical joint program and a strategic alliance would consist of collaboration between different stagnating industries. Here network operators, network manufacturers and media houses, which are strategic relatives, could join their forces in a new way to build unforeseen models and mechanisms to monetarise content creation, online publishing, vlogging and blogging, developer communities, editing, visualisation, internet radio and TV, virtual and augmented reality, Fig. 1. The TAIC-SIMO tetra model shows the fundamental relation between four different industries in (e.g. network operator–network manufacturer–media–Internet). They are all dependent of each other so that the media serves content (Interest) to the Internet whereas the network operator (Channel) and manufacturer (Technology) enable the Internet services on the background. The Access is the user's favorite home entry page to the Internet [25].

A third avenue, to "Uberise" your corporation, is not explained but only mentioned in this paper. The idea of a completely disruptive model for an industry usually comes from startups e.g. Uber to offer taxi services, or a disruption to another industry as by-product e.g. Google was originally not a media company, but its search absorbed journalism's income, Google took journalism's paycheck, as we call it.

Like Bontemps says, uberisation of business is an increase in volume and expansion through several new and different fields. [26]. In large corporations uberisation could happen either through making a radical change in their business focus, products, services and leadership, or through trying to change the surrounding business environment and markets. Usually these actions are impossible because of the magnitude and investments of their present business and because of corporative ownership and management. The needed radical decisions to be made hardly get support amongst the shareholders, investors and board members. Only radical startups change the business environment on conventional and ordinary industries and change happens insidiously, like changing the media through search and chat.

## V. Conclusions and Aftermath

Without radical actions journalism does not survive in the pressure of technological development.

The circumstances of media and journalism are more uncertain and unpredictable. There are more questions, opinions and claims than innovation, radical actions and academic or engineered solutions for gaining back media's and journalism's grandeur. Software startuppers and their technological developments are driving tremendous change in media and journalism, which both have major societal roles. Journalists, political scientists and philosophers are merely trying to keep up the pace and observing the change, not leading it themselves. But shouldn't they be doing exactly that? Why don't they?

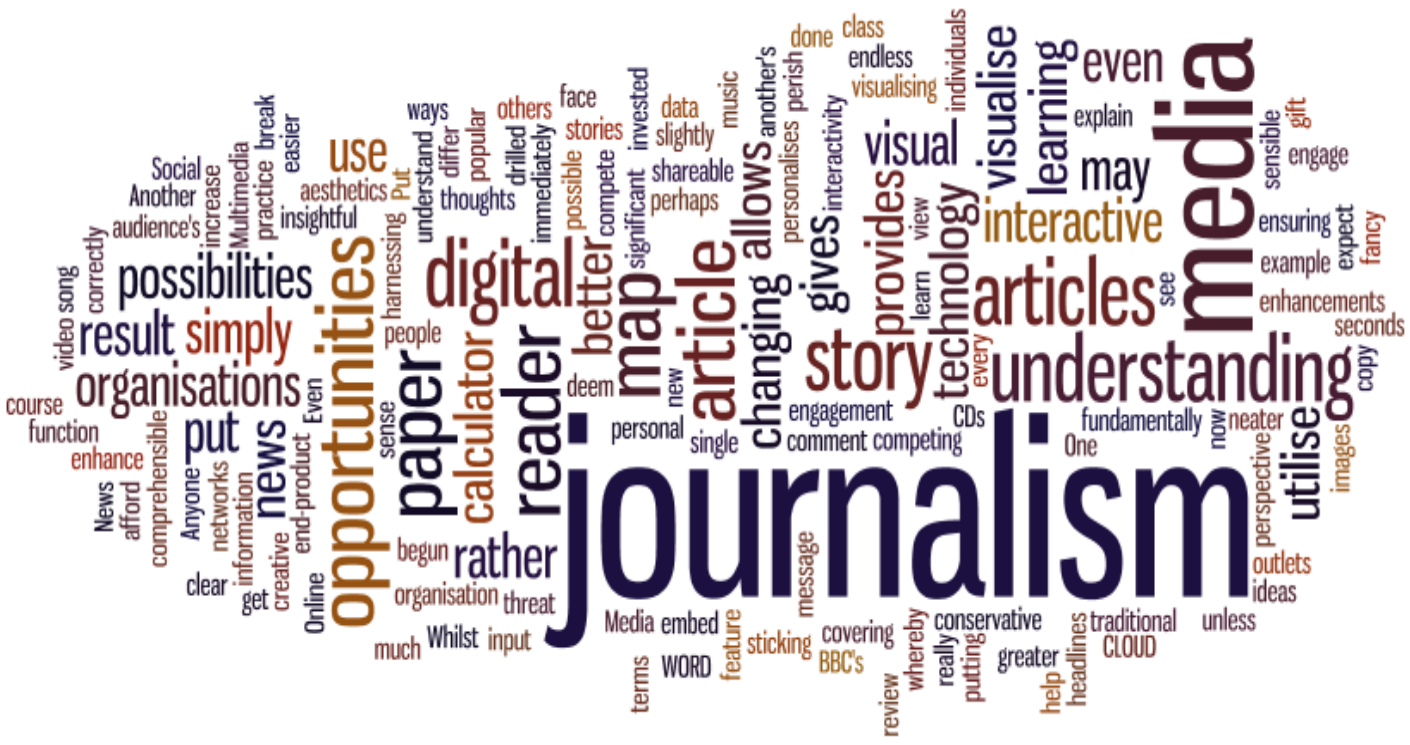

Fig. 4. Present day environment of journalism is filled with uncertainty, possibilies, jargon and gradual only actions. The picture above represents the status quo in many ways: there are 150 notions related to journalism, all the notions are either opportunities or reasons for further confusion. Perhaps more descriptive in the picture is that it is hard to find its origins: two different journalistic notions quote to each other. [27], [28].

If startuppers killed media and journalism – is the Fourth Estate dead? Who's the watchdog now? And should bloggers and other citizen journalists be granted access to press conferences and other sources of information traditionally reserved for those holding an official press card? What does journalism cost and why – and who should pay? These are few of the questions for the philosophers.

Traditional media relies on getting its information mostly from official sources (e.g. official press releases and conferences, press agencies) and through conducting personal interviews. It is thus merely repeating what an authority or a person said. Is that still the right way of doing it? Do traditional journalistic ways of working need redefinition? What is quality and how is political correctness linked to quality? These are few of the questions to the media professionals and academics.

From a systemic operator-manufacturer-media house collaboration totally new products, services, patents, formats and processes can be developed. The operator-manufacturer-media house Trio finds new startups and developers through establishing a global developer and creative community of hundreds of millions of people all equipped with smartphones, action cameras, electric bikes, editing software, bloggers' creativity and new creative culture.

Even a car with its in-car and car-to-car communication can be developed into a car reporter and journalist, which would be a radical abstraction shift. The car becomes anyway a scanner, camera and a total sensor, so a journalistic platform with technologies and formats for media houses could well be an equivalent of Google's Street View magnitude. This would be business as usual for the network manufacturer. This assumption suggests a window of opportunity for disruptive car-to-car innovation for informing, connecting and entertaining the drivers, passengers and citizens in cities and motorways as they read, listen and watch content on the road. The network manufacturer could easily build technology to manage all data flows, payments, videos, chats, studios, media centers and creative reporters' equipment. The network operator could easily build a local, national and continental delivery system on top of the manufacturer's and media house's innovations. Here three industries filled with uncertainty would work together with the ethics of journalism, manufacturer's engineering skills and operator's local customer base.

Operator-manufacturer-media house Trio finds easily strategic partners from areas where they don't have harsh competitors. The idea is that the Trio as an Internet veteran and trusted content creator takes the customer, user and creator as a partner, pays properly for the content, promotes the creative reporter community and acts like a peer startupper.

Industry evolution on the Trio's three specific, business areas need research, development and discussion, both singular and cross-sectoral. Research in the rapid evolution and changes in media industry, startup culture and technology has its inertias of research practices to reach applicable results in time. Therefore research should always be a counterpart of development. Development for its part happens either by the slow and at times pompous corporations or by novice startuppers with more eagerness than sense of professional execution. When Giardino (et.al.) [29] refer to Marmer (et.al.) [30] that more than 90 % of startups fail, no industry, trade or human activity can accept this enormous waste of startuppers' work. This is both a research and development question.

The Trio's fightback happens best through repairing journalism and making these industries become part of a highly ethical commitment. Here the participation of the faculties of social sciences and humanities is crucial. Software startuppers need partners from those areas to avoid a situation in which the societal development is allowed to happen "technology first". Ethical and societal startups need to stem from the faculties of drama, journalism, philosophy, social and political sciences to become part of and influence the rapid development in businesses that software startuppers have alone boosted.

## Acknowledgment

This paper has been a hard effort to formulate a proposal for several different industries to join their forces with startup culture. Pekka Abrahamsson has supported all hard initiatives which could be called complex systemic formulations. But, in his words, "write and produce text first, otherwise nothing happens." We want to thank Pekka for his encouraging words.

## References

Beside scientific sources, media and journalistic articles and news have been taken as sources for references because the biggest concern about media's and journalism's economical downfall is expressed immediately on their news and articles. Media and journalism also try to find solutions for the fightback to get the advertisement money from Web services like Google, Facebook and Twitter.

As a source, media and journalism should be trustworthy because of their self-regulation, peer feedback, and especially because of their ethical rules and publishing policies. This applies, according to the so called quality media, at least to the quoted references. The reliability of the media and journalism is based on the idea of freedom of speech.

One important reason to have media and journalism as a source of data is that media and journalism trust on their own surveys, analyses and action models and they act according to these findings.